\global\def\draftcontrol{0}

   \def\versionno{holreheat}
\catcode`\@=11
\expandafter\ifx\csname draftcontrol\endcsname\relax\global\def\draftcontrol{0}
\fi

{\count255=\time\divide\count255 by 60
\xdef\hourmin{\number\count255}
\multiply\count255 by-60\advance\count255 by\time
\xdef\hourmin{\hourmin:\ifnum\count255<10 0\fi\the\count255}}
\def\draftdate{\number\month/\number\day/\number\year\ \ \ \hourmin }

\newcommand\makepapertitle{\par
  \begingroup
    \renewcommand\thefootnote{\@fnsymbol\c@footnote}%
    \def\@makefnmark{\rlap{\@textsuperscript{\normalfont\@thefnmark}}}%
    \long\def\@makefntext##1{\parindent 1em\noindent
            \hb@xt@1.8em{%
                \hss\@textsuperscript{\normalfont\@thefnmark}}##1}%
     \newpage
     \global\@topnum\z@   % Prevents figures from going at top of page.
     \@makepapertitle
     \thispagestyle{empty}\@thanks
  \endgroup
  \setcounter{footnote}{0}%
  \global\let\thanks\relax
  \global\let\makepapertitle\relax
  \global\let\@makepapertitle\relax
  \global\let\@thanks\@empty
  \global\let\@author\@empty
  \global\let\@date\@empty
  \global\let\@title\@empty
  \global\let\title\relax
  \global\let\author\relax
  \global\let\date\relax
  \global\let\and\relax
  \def\version{\let\version\@version\@gobble}
}
\def\@makepapertitle{%
  \newpage
   \ifnum\draftcontrol=1 {}
   \version\versionno
   \vskip 3em%
   \else
   \hfill\hbox to 3cm {\parbox{4cm}{\@pubnum}\hss}%
   \vskip 3em%
   \fi
   \begin{center}%
   \let \footnote \thanks
     {\LARGE {\@title}}%
     \vskip 1.5em%
     {\normalsize%\large
       \lineskip .5em%
       \begin{tabular}[t]{c}%
         \@author
       \end{tabular}\par}%
     \vskip 1.5em%
     {\@bstract}%
     \end{center}%
     \vskip 1.5em
     \@date%
   \par
}

\gdef\@pubnum{}
\def\pubnum#1{%
  \gdef\@pubnum{#1}}

\gdef\@bstract{}
\def\Abstract#1{%
  \gdef\@bstract{%
   \parbox{\textwidth-0pc}{%
   \centerline{\bf Abstract}\penalty1000%
\kern.2cm%
\noindent%\abstractfont \baselineskip=12pt
\renewcommand\baselinestretch{1.0}%
{#1}}}
}

\def\ps@paper{\let\@mkboth\@gobbletwo%
     \ifnum\draftcontrol=1
    \def\@oddfoot{\hbox to \textwidth{\tiny \versionno \hfil\tiny\draftdate}%
    \hskip -\textwidth \hbox to \textwidth{\hfil\rm\thepage\hfil}}%
     \else\def\@oddfoot{\hbox to \textwidth{\hfil\rm\thepage\hfil}}
     \fi
     \let\@evenfoot\@oddfoot
}
\def\body{\clearpage
          \pagestyle{paper}
    }
\def\@version#1{\ifnum\draftcontrol=1
\typeout{}\typeout{#1}\typeout{}
\vskip3mm\centerline{\hbox{\fbox{\normalsize{\tt DRAFT -- #1 -- }
                   {\draftdate}}}}\vskip3mm
\fi}
\let\version\@version
\long\def\eqlabel#1{\ifnum\draftcontrol=1
                    \tag@false  % there are some problems with multline without this
                    \tag*{(\theequation) \hbox to -0.2cm{\hspace{0cm}\small{#1}\hss}}
                    \refstepcounter{equation}
                    \edef\@currentlabel{\theequation}
                    \ltx@label{#1}          % use old LaTeX \label instead of new definition
                    \else
                    \label{#1}
                    \fi
                    }
\let\st@bibitem\@bibitem
\let\st@lbibitem\@lbibitem
\ifnum\draftcontrol=1
  \def\@bibitem#1{%
    \st@bibitem{#1}\a@@label{#1}\ignorespaces}
  \def\@lbibitem[#1]#2{%
    \st@lbibitem[#1]{#2}\a@@label{#2}\ignorespaces}
  \def\a@@label#1{%
    \gdef\a@lab{\smash{\normalfont\small#1}}
    \ifvmode
      \if@inlabel
        \global\setbox\@labels\hbox{%
          \llap{\a@lab\let\a@lab\relax
                \kern\@totalleftmargin\kern\marginparsep}%
          \box\@labels}%
      \fi
    \fi}
\fi
\documentclass[12pt,letterpaper]{article}
\usepackage{amsmath,amssymb,array,calc,epsfig,rotating,bm,xcolor}
\usepackage[sort]{cite}
\usepackage{graphicx,esint,float}
\usepackage{psfrag,verbatim}
\usepackage[makeroom]{cancel}
\usepackage{xcolor,url}
\usepackage{hyperref}
\ifnum\draftcontrol=1
\fi
\renewcommand\baselinestretch{1.25}
\renewcommand\section{\@startsection {section}{1}{\z@}%
                                   {-3.5ex \@plus -1ex \@minus -.2ex}%
                                   {2.3ex \@plus.2ex}%
                                   {\normalfont\large\bfseries}}
\renewcommand\subsection{\@startsection{subsection}{2}{\z@}%
                                   {-3.25ex\@plus -1ex \@minus -.2ex}%
                                   {1.5ex \@plus .2ex}%
                                   {\normalfont\normalsize\bfseries}}
\renewcommand\subsubsection{\@startsection{subsubsection}{3}{\z@}%
                                   {-3.25ex\@plus -1ex \@minus -.2ex}%
                                   {1.5ex \@plus .2ex}%
                                   {\normalfont\normalsize\it}}
\renewcommand\paragraph{\@startsection{paragraph}{4}{\z@}%
                                   {-3.25ex\@plus -1ex \@minus -.2ex}%
                                   {1.5ex \@plus .2ex}%
                                   {\normalfont\normalsize\bf}}

\numberwithin{equation}{section}

\def\revise#1       {\raisebox{-0em}{\rule{3pt}{1em}}%
                     \marginpar{\raisebox{.5em}{\vrule width3pt\
                     \vrule width0pt height 0pt depth0.5em
                     \hbox to 0cm{\hspace{0cm}{%
                     \parbox[t]{4em}{\raggedright\footnotesize{#1}}}\hss}}}}

\newcommand\nxt[1]  {\\\fnxt#1}
\newcommand{\ie}{{\it i.e.,}\ }
\newcommand{\eg}{{\it e.g.,}\ }

\def\cale         {{\cal E}}
\def\calf         {{\cal F}}

\def\calh         {{\cal H}}

\def\call         {{\cal L}}
\def\calm         {{\cal M}}
\def\caln         {{\cal N}}
\def\calo         {{\cal O}}

\def\cals         {{\cal S}}

\def\del          {\partial}

\def\sqr#1#2{{\vcenter{\vbox{\hrule height.#2pt
 \hbox{\vrule width.#2pt height#1pt \kern#1pt
 \vrule width.#2pt}\hrule height.#2pt}}}}

\def\aa1{\phi}
\def\cc1{\psi}

\def\f0{\text{\boldmath$\varphi$}}
\def\h2{\mathfrak{h}}

\def\qft{{\rm QFT}_{d+1}}
\def\cft{{\rm CFT}_{d+1}}

\catcode`\@=12

\begin{document}

%%%
%%%%%% text starts here
%%%%%%%%%

\title{%
  {\bf Gravitational reheating at strong coupling}}

\date{April 21, 2023}
%\date\today

\author{
Alex Buchel\\[0.4cm]
\it $ $Department of Physics and Astronomy\\ 
\it University of Western Ontario\\
\it London, Ontario N6A 5B7, Canada\\
\it $ $Perimeter Institute for Theoretical Physics\\
\it Waterloo, Ontario N2J 2W9, Canada
}

\Abstract{We use the gauge/gravity correspondence to study the gravitational
reheating of strongly coupled gauge theories in the rapid exit from
the long inflationary (de Sitter) phase.  We estimate the maximal
reheating temperature of holographic models in $d$-spatial dimensions,
whose scale invariance is explicitly broken by a source term for an
operator ${\cal O}_\Delta$ of dimension $\frac d2+\frac 12< \Delta< d+1$,
as well as the top-down ${\cal N}=2^*$ and the cascading gauge theories.
The reheating is most efficient when the inflationary phase Hubble
constant is much larger than the mass scale of the conformal symmetry
breaking of the theory, and for theories with the conformal symmetry
breaking operators that are close to marginal.
}

\makepapertitle

\body

\version\versionno
%\tableofcontents

\section{Introduction and summary}
\label{intro}

Gauge theory/string theory correspondence \cite{Maldacena:1997re,Aharony:1999ti}
is a valuable tool to study dynamics of strongly coupled gauge theories, where
alternatives are scarce to non-existing. It served as an inspiration
of novel insight into quantum field theory non-equilibrium dynamics, as, \eg in
\cite{Baier:2007ix,Bhattacharyya:2007vjd,Chesler:2008hg}.

Holography was instrumental in discovery of de Sitter dynamical fixed points (DFPs) of a
Quantum Field Theory \cite{Buchel:2017pto,Buchel:2021ihu}. Specifically,
consider $d+1$-spacetime dimensional quantum field theory, $\qft$, in
Friedmann-Lemaitre-Robertson-Walker (FLRW) Universe with a scale factor $a(t)$,
\begin{equation}
ds_{d+1}^2=-dt^2+a(t)^2\ d\bm{x}^2\,.
\eqlabel{at}
\end{equation}
Assuming that the FLRW Universe is asymptotically de Sitter,
\begin{equation}
\lim_{t\to +\infty}\frac{\dot a}{a}=H>0\,,
\eqlabel{dsass}
\end{equation}
an arbitrary initial state of the $\qft$ would typically evolve to a de Sitter DFP ---
an internal state of the theory with spatially homogeneous and time-independent one-point
correlation functions of its stress-energy tensor $T^{\mu\nu}$, and a set of gauge-invariant
local operators $\{\calo_i\}$, along with strictly positive divergence of the entropy
current $\cals^\mu$ at late times,
\begin{equation}
\lim_{t\to\infty}\biggl(\nabla\cdot \cals\biggr)\ >\ 0\,.
\eqlabel{divs}
\end{equation}
The everlasting entropy production implied by \eqref{divs} indicates
that a DFP is a genuinely non-equilibrium state of the QFT. A given theory might
have multiple distinct DFPs, characterized by a local order parameter of a
spontaneously broken global symmetry; there can be phase transitions between DFPs
--- these and other aspects of a de Sitter DFP of a quantum field theory with a holographic
dual were explored in \cite{Buchel:2016cbj,Buchel:2017pto,Buchel:2017qwd,
Buchel:2017lhu,Buchel:2019qcq,Buchel:2019pjb,Casalderrey-Solana:2020vls,Buchel:2021ihu,
Ecker:2021cvz,Penin:2021sry,Bea:2021zol,Buchel:2022hjz,Buchel:2022sfx}.

To understand the astrophysical consequence of a DFP it is useful to unpack \eqref{divs}.
Assuming spatial homogeneity and isotropy, we associate the entropy current $\cals^\mu$ to
a comoving observer $u^\mu\equiv (1,\bm{0})$ as $\cals^\mu=s(t) u^\mu $, where $s(t)$
is the physical entropy density. Then,
\begin{equation}
\lim_{t\to \infty} \biggl(\nabla\cdot S\biggr)
=\lim_{t\to\infty}\frac{1}{a(t)^d}\ \frac{d}{dt}\left(a(t)^d s(t)\right)=
d\ H\ s_{ent}\,,
\eqlabel{ns2}
\end{equation}
where the late-time, $t\to \infty$, limit of the physical entropy density $s(t)$, provided this
limit exists, was called in \cite{Buchel:2017qwd} the  {\it vacuum entanglement entropy} (VEE)
density
\begin{equation}
\lim_{t\to\infty}s(t)=s_{ent}\,.
\eqlabel{slim}
\end{equation}
In other words, asymptotically finite entropy production rate of a QFT de Sitter DFP is
due to asymptotically finite physical entropy density. The latter point dramatically
differentiates conformal and non-conformal theories: for a CFT it is not the physical,
but rather the comoving
entropy density $s_{comoving}(t)\equiv a(t)^d s(t)$ that is asymptotically
constant\footnote{See section 2 of \cite{Buchel:2019qcq}
for a pedagogical holographic review.}
in de Sitter space-time; the physical entropy density of a CFT rather vanishes
in asymptotically de Sitter space-time. As a result, the entropy production rate for a CFT in asymptotically
de Sitter space-time vanishes as well --- a CFT can not have de Sitter
DFPs\footnote{One can view this as  a reflection
of the fact that a CFT dynamics in de Sitter is Weyl equivalent
to a dynamics in Minkowski space-time; given that Minkowski dynamics leads to equilibration,
the same must be true for a CFT late-time de Sitter evolution.
}. Consider now a realistic cosmology where an extended period of the accelerated expansion
(the inflationary de Sitter phase) is followed by the inflationary exit, and subsequently the
Hot Big Bang. In \cite{Buchel:2023djl} it was argued that the Hot Big Bang (HBB) is a
{\it generic} outcome of the fast exit from the de Sitter DFP, irrespectively
of the inflaton physics\footnote{In particular,
whether or not the inflaton couples directly to the
theory to be reheated.}. The argument is very simple, and we review it here
for completeness:
\begin{itemize}
\item While technical details of the inflationary exit will not be important,
it is useful to keep in mind a specific model. To this end we assume that
the Hubble parameter $\calh$ evolves as
\begin{equation}
\calh(t)\equiv \frac{d}{dt}\ln a(t)=\frac{H}{1+\exp(2\gamma t)}\,.%\qquad
%a(t)=\left(\frac{2}{1+\exp(-2\gamma t)}\right)^{\frac{H}{2\gamma}}\,.
\eqlabel{exita}
\end{equation}
Here the constant energy scale $\gamma$ specifies the exit-rate from
the inflation, 
and the scale parameter $a(t)$ is normalized as $a(t=0)=1$.
For a rough estimate, in this model the exit from inflation
occurs during the time frame
$t\in\  \propto (-\gamma^{-1},\gamma^{-1})$, so that the scale factor
changes as
\begin{equation}
\ln a(t)\bigg|_{start}^{end}\equiv \ln\frac{a_e}{a_s}\ \sim\
\int_{-{\gamma^{-1}}}^{\gamma^{-1}} dt\
\frac{H}{1+\exp(2\gamma t)}=\frac{H}{\gamma}\,.
\label{dela}
\end{equation}
\item During the inflationary exit, the comoving entropy density
can only increase, so
\begin{equation}
\underbrace{s_{comoving,s}}_{\sim a_s^{d}\cdot s_{ent}}\ \le\
\underbrace{s_{comoving,e}}_{\equiv a_e^{d}\cdot s_e} \,,
\eqlabel{dels}
\end{equation}
where we denoted by $s_e$ the 
physical entropy density of the post-inflationary QFT state. 
\item From \eqref{dels} we conclude that
\begin{equation}
s_e\ \sim\ s_{ent}\ \left(\frac{a_s}{a_e}\right)^{d}\ \sim s_{ent}\cdot
e^{-\frac{Hd}{\gamma}} = s_{ent}\cdot \calo(1)\,,
\end{equation}
with $\calo(1)\to 1$ in the fast exit limit $\gamma\gg H$.
\item The post-inflationary state $ _e$ is nonequilibrium; its subsequent evolution
leads to its thermalization with the Hot Big Bang thermal entropy density
\begin{equation}
s_{HBB} \ge s_e\ \ge s_{ent} \,.
\eqlabel{shbb}
\end{equation}
\item To put a lower bound on the
maximal HBB reheating temperature $T_{HBB}$, one can interpret the
VEE density $s_{ent}$ of a theory de Sitter DFP as an equilibrium thermal
entropy density of the theory $s_{eq}(T)$ at the {\it DFP effective temperature}, $T_{DFP}$,
\begin{equation}
T_{HBB}\ \ge T_{DFP}\,,\qquad {\rm where}\qquad s_{eq}(T)\bigg|_{T=T_{DFP}}=s_{ent}\,.
\eqlabel{lbt}
\end{equation}
\end{itemize}
In this paper
we strengthen the  claim of \cite{Buchel:2023djl}, in particular we argue that the reheating is
very efficient, provided the inflationary stage Hubble scale $H$ is much higher
than any mass scale $\Lambda$ of the theory to be reheated, \ie when
\begin{equation}
\frac{H}{\Lambda}\gg 1 \,.
\eqlabel{hl}
\end{equation}

We now summarize our results:
\nxt Consider a massive $\qft$ that is a deformation of a CFT$ _{d+1}$ by a
single operator $\calo_\Delta$ of
dimension $\frac d2 +\frac 12 <\Delta< d+1$,
\begin{equation}
\call_{QFT}=\call_{CFT}+\Lambda^{d+1-\Delta}\ \calo_\Delta\,.
\eqlabel{lqft}
\end{equation}
To the leading order in $\calo\left(\frac{\Lambda}{H}\right)$ we find
\begin{equation}
\begin{split}
&\frac{T_{DFP}}{\Lambda}=\left[\ \calf(d,\Delta)+\calo\left(\left(\frac{\Lambda}{H}
\right)^{\frac{2(d+1-\Delta)}{d+1}}\right)\
\right]\cdot \biggl(\frac{H}{\Lambda}\biggr)^{\frac{2 \Delta-d-1}{d+1}}\ \gg 1\,,
\end{split}
\eqlabel{ltdfp}
\end{equation}
where
\begin{equation}
\calf(d,\Delta)=\frac{d+1}{4\pi}  \left(
\frac{2^{d- 2\Delta} \pi^3 d}{(d+1)
\Gamma\left(\Delta-\frac d2-\frac12\right)^2 \Gamma\left(1-\frac d2\right)^2 \Gamma
\left(d+1-\Delta\right)^2
\sin^2\frac{\pi d}{2} }
\right)^{\frac{1}{d+1}}\,.
\eqlabel{deff}
\end{equation}
Following \eqref{lbt}, the reheating temperature $T_{HBB}$ is at least as high
as that of $T_{DFP}$ in \eqref{ltdfp}, \ie the reheating is very efficient
in the phenomenologically relevant regime $H\gg \Lambda$.
Note that
QFTs with operators of  higher dimensions achieve larger
reheating temperature relative to the mass scale of the theory; however
$\calo_\Delta$ can not be marginal --- from \eqref{deff},
\begin{equation}
\calf(d,\Delta)\bigg|_{\Delta=d+1}=0\,.
\eqlabel{fmag}
\end{equation}
In fact, we expect that for an exactly marginal operator $\calo_{d+1}$
the reheating temperature exactly vanishes, since the theory remains conformal
in this case.
\nxt Typically a massive QFT has nonzero sources for a set of relevant operators
$\{\calo_{\Delta_i}\}$. For example, $\caln=2^*$ gauge theory
\cite{Pilch:2000fu,Buchel:2000cn,Buchel:2013id} is a mass-deformation
of $\caln=4$ supersymmetric Yang-Mills theory by a pair of operators
$\{\calo_2,\calo_3\}$. We show that in such cases the maximal reheating temperature
is still bounded from below by \eqref{ltdfp}, where $\Delta=\max\{\Delta_i\}$, and $\Lambda$
is the mass scale of the source of the maximal dimension operator in the range
$\frac d2+\frac 12<\Delta<d+1$.
\nxt The Klebanov-Strassler cascading gauge theory \cite{Klebanov:2000hb,Herzog:2001xk}
does not have relevant operators; it does have a single marginal, but not exactly marginal,
operator \cite{Aharony:2005zr}. In this case we find that the
maximal reheating temperature
is bounded from below by $T_{DFP}$ with  
\begin{equation}
\frac{T_{DFP}}{\Lambda}\ \sim\ \frac{H}{\Lambda} \cdot\frac{1}{\left(\ln
\frac{H}{\Lambda}\right)^{11/12} }\ \gg 1\,,
\eqlabel{tdfpkt}
\end{equation}
where $\Lambda$ is the strong coupling scale of the confining cascading gauge
theory\footnote{See section 2 of \cite{Buchel:2021yay}.}.

The rest of the paper is organized as follows.
In section \ref{perd} we derive \eqref{ltdfp}.
In section \ref{n2} we combine the earlier results
on $\caln=2^*$ de Sitter DFPs \cite{Buchel:2017pto}
and its Minkowski space-time thermodynamics \cite{Buchel:2007vy}
to derive the corresponding $T_{DFP}$. In section
\ref{ksr} we combine the  earlier
results on the cascading gauge theory de Sitter DFPs
\cite{Buchel:2019pjb} and its high-temperature
Minkowski space-time thermodynamics \cite{Buchel:2000ch,Aharony:2007vg}
to derive the corresponding $T_{DFP}$ \eqref{tdfpkt}.

The reheating mechanism introduced in \cite{Buchel:2023djl} and further developed here
is a strong coupling (holographic) version of the preheating in "non-oscillatory'' (NO) models 
\cite{Peebles:1998qn,Felder:1999pv}. To make a closer connection it would be
interesting to study finite-$N$ and finite 't Hooft coupling corrections to
the holographic gravitational reheating discussed here.

\section{Efficient near-conformal reheating in $\qft$}
\label{perd}

Consider a holographic toy model of $d+1$-dimensional massive $\qft$ with the effective dual
gravitational action\footnote{We set the radius $L$ of an asymptotic
$AdS_{d+2}$ geometry to unity.}:
\begin{equation}
S_{d+2}=\frac{1}{2\kappa^2}\int_{\calm_{d+2}}d^{d+2}x\sqrt{-g}\biggl[R+d(d+1)-\frac 12
\left(\nabla\phi\right)^2-\frac{m^2}{2}\phi^2\biggr]\,.
\eqlabel{seffd}
\end{equation}
The $d+2$-dimensional gravitational constant $\kappa$ is related to the
ultraviolet (UV) conformal fixed point $\cft$  central charge,
and 
$\phi$ is  a  gravitational  bulk scalar with 
\begin{equation}
L^2 m^2=\Delta(\Delta-d-1) \;,
\eqlabel{mphi}
\end{equation}
which is dual to a dimension $\Delta$ operator $\calo_\Delta$ of the boundary theory. 
$\qft$ is a relevant deformation of the UV  $\cft$ as in \eqref{lqft}.
We study $\qft$ dynamics 
 in FLRW Universe \eqref{at}.

\subsection{Holographic gravitational dynamics of $\qft$}

A generic state of the boundary field theory with a gravitational dual \eqref{seffd}, homogeneous and isotropic in the spatial
boundary coordinates $\boldsymbol{x}=\{x_1,\cdots,x_d\}$, leads to a bulk gravitational metric ansatz
\begin{equation}
ds_{d+2}^2=2 dt\ (dr -A dt) +\Sigma^2\ d\boldsymbol{x}^2\,,
\eqlabel{EFmetric}
\end{equation}
with the warp factors $A,\Sigma$ as well as the bulk scalar $\phi$
depending only on $\{t,r\}$. From
the effective action \eqref{seffd} we obtain the following equations of
motion:
\begin{equation}
\begin{split}
&0=d_+'\Sigma+(d-1)\ d_+\Sigma\ \left(\ln\Sigma\right)'+\frac{\Sigma}{4d}\ (m^2\phi^2 -2d(d+1))\,,\\
&0=d_+'\phi+\frac d2\ d_+\phi\ \left(\ln\Sigma\right)'+\frac d2\ \frac{d_+\Sigma}{\Sigma}\ \phi'
-\frac{m^2}{2}\ \phi\;,\\
&0=A''-d(d-1)\ \frac{d_+\Sigma}{\Sigma^2}\ \Sigma'+\frac 12 d_+\phi\ \phi'
-\frac{m^2(d-2)}{4d}\ \phi^2+\frac12 (d+1)(d-2)\,,
\end{split}
\eqlabel{ev1}
\end{equation}
as well as the Hamiltonian constraint equation:
\begin{equation}
0=\Sigma''+\frac{1}{2d}\  (\phi')^2\ \Sigma\,,
\eqlabel{ham}
\end{equation}
and the momentum constraint equation:
\begin{equation}
\begin{split}
&0=d_+^2\Sigma-A'\ d_+\Sigma+\frac{1}{2d}\ \Sigma\ (d_+\phi)^2\,.
\end{split}
\eqlabel{mom}
\end{equation}
In \eqref{ev1}-\eqref{mom} and below  
we denoted $'= \frac{\del}{\del r}$, $\dot\ =\frac{\del}{\del t}$, 
and $d_+= \frac{\del}{\del t}+A \frac{\del }{\del r}$. 
The near-boundary $r\to\infty$ asymptotic behavior
of the metric
functions and the scalar encode the mass parameter $\Lambda$ and the boundary
metric scale factor $a(t)$:
\begin{equation}
\begin{split}
&\Sigma=a\biggl({r}+\lambda+\cdots\biggr)\,,\qquad A=\frac{r^2}{2}
+\left(\lambda-\frac{\dot a }{a }\right)r+\cdots\,,
\qquad \phi=\left(\frac{\Lambda}{r}\right)^{d+1-\Delta}+\cdots\,.
\end{split}
\eqlabel{bcdata}
\end{equation}
$\lambda=\lambda(t)$ in \eqref{bcdata} is the residual radial coordinate diffeomorphism parameter \cite{Chesler:2013lia}.
An initial state of the boundary field theory is specified providing the scalar
profile $\phi(0,r)$ and solving the
constraint \eqref{ham}, subject to the boundary
conditions \eqref{bcdata}. Equations \eqref{ev1} can then be used to evolve
the state.

The subleading terms in the boundary expansion of the
metric functions and the scalar encode the evolution of the  energy
density $\cale(t)$, the pressure $P(t)$ and the expectation values of the operator
$\calo_\Delta(t)$ of the prescribed boundary QFT initial state. These observables
can be computed following the holographic renormalization of the model as, \eg for models
discussed in \cite{Buchel:2017pto} and \cite{Buchel:2017lhu}. Our  interest here is
however the dynamics of the non-equilibrium entropy density $s(t)$. 
The background metric \eqref{EFmetric} 
has an apparent horizon located at $r=r_{AH}$, where \cite{Chesler:2013lia}
\begin{equation}
d_+\Sigma\bigg|_{r=r_{AH}}=0\,.
\eqlabel{defhorloc}
\end{equation} 
Following \cite{Booth:2005qc,Figueras:2009iu} we associate the non-equilibrium  entropy density $s$
of the boundary $\qft$  with the Bekenstein-Hawking entropy density of the apparent horizon  
\begin{equation}
s_{comoving}= a^d s =\frac {2\pi}{\kappa^2}\ {\Sigma^d}\bigg|_{r=r_{AH}}\,.
\eqlabel{as}
\end{equation}
Using the holographic background equations of motion \eqref{ev1}-\eqref{mom} 
we find 
\begin{equation}
\frac{d(s_{comoving})}{dt}= \frac{d(a^d s)}{dt}=\frac{4\pi}{\kappa^2}\ (\Sigma^d)'\ \frac{
 (d_+\phi)^2}{\Delta(d+1-\Delta)\phi^2+2d(d+1)}\bigg|_{r=r_{AH}}\,.
\eqlabel{dasdt}
\end{equation}
From the Hamiltonian constraint \eqref{ham} and the boundary asymptotic \eqref{bcdata},
\begin{equation}
\Sigma'(t,r)=a(t)+\frac{1}{2d}\int_r^\infty d\rho\ \Sigma(t,\rho)\ \left(\frac{\del\phi(t,\rho)}{\del\rho}\right)^2
\ >\ 0\,,
\eqlabel{possigma}
\end{equation}
\ie is manifestly positive; thus we conclude that during the holographic evolution, the comoving entropy
density can only increase 
\begin{equation}
\frac{d(s_{comoving})}{dt}\ge 0\,.
\eqlabel{dasdt2}
\end{equation}
The latter establishes the vital fact in harvesting the vacuum entanglement entropy of a de Sitter DFP,
see \eqref{dels} and \eqref{shbb}.

\subsection{de Sitter DFP of $\qft$}

Following \cite{Buchel:2017pto}, the equations for the late-time attractor of the evolution (a de Sitter DFP)
can be obtained from \eqref{ev1}-\eqref{mom} taking $t\to \infty $ limit with identification
\begin{equation}
\lim_{t\to\infty}\{\phi,A\}(t,r)=\{\phi,A\}_v\,,\qquad \lim_{t\to\infty} \frac{\Sigma(t,r)}{a(t)}=\sigma_v(r)\,.
\eqlabel{ds3vac}
\end{equation}
Introducing a new radial coordinate 
\begin{equation}
x\equiv \frac Hr\,,
\eqlabel{defx}
\end{equation}
and denoting 
\begin{equation}
A_v=\frac{H^2}{2x^2}\ g(x)\,,\qquad \sigma_v=\frac {H}{x}\ f(x)\,,
\eqlabel{redefstat}
\end{equation}
we find 
\begin{equation}
\begin{split}
&0=f''+\frac {1}{2d}\ (\phi_v')^2\ f\,,\\
&0=\phi_v''-\biggl(
f x^2 (f \phi_v^2 m^2-4 d x^2 (d+1) f'+2 f d (2 x-1) (d+1))
\biggr)^{-1}
\biggl(
f^2 x^4 (d+2) (\phi_v')^3\\
&+\phi_v f^2 x^2 m^2 (\phi_v')^2+\phi_v' (2 d x^4 (d+2)
(d+1) (f')^2-f x^2 (\phi_v^2 m^2 (d+2)+2 d (d+1) (2 d x\\
&-d-2)) f'+f^2 d x (m^2 \phi_v^2+(2 (d+1)) (d x-d-2 x)))
-2 \phi_v d m^2 (f' x-f)^2 (d+1)
\biggr)\,,
\end{split}
\eqlabel{fpeoms}
\end{equation}
along with an algebraic expression for $g$:
\begin{equation}
g=\frac{f (f \phi_v^2 m^2-2 d (d+1) (2 f' x^2-f (2 x-1)))}
{f^2 (\phi_v')^2 x^2-2 d (f' x-f)^2(d+1)}\,.
\eqlabel{geom}
\end{equation}
A de Sitter DFP solution has to satisfy the boundary conditions \eqref{bcdata}, and remain nonsingular 
for $x\in (0,x_{AH}]$, where the location of the apparent horizon $x_{AH}$ is determined from 
\cite{Buchel:2017pto}
\begin{equation}
d_+\Sigma(t, x_{AH})=0\qquad \Longleftrightarrow\qquad \biggl(f(x)\ (2 x +g(x)) -x g(x) f'(x)\biggr)\bigg|_{x=x_{AH}}=0\,.
\eqlabel{xah}
\end{equation}
 Without the loss of generality 
we fix the diffeomorphism parameter $\lambda$ so that 
\begin{equation}
A_v(x)\bigg|_{x=\frac 13}=0\,.
\eqlabel{lstat}
\end{equation}
We will always have $x_{AH}>\frac 13$.

For general values of $\frac{\Lambda}{H}$,  solutions representing  de Sitter DFPs
have to be found numerically. Luckily, the regime of the efficient reheating
happens when the conformal symmetry breaking parameter 
\begin{equation}
p_1\equiv \left(\frac{\Lambda}{H}\right)^{d+1-\Delta}
\eqlabel{defp1}
\end{equation} 
is small; in the latter case the solution to \eqref{fpeoms} can be constructed
perturbatively in $p_1$. To proceed, we set
\begin{equation}
\begin{split}
&\phi_v=p_1\ p_{(1)}(x)+\calo(p_1^3)\,,\qquad f=(1-x)\left(1+p_1^2\ s_{(2)}(x)+\calo(p_1^4)\right)\,,\\
&g=(1-3x) \left(1-x+2p_1^2x\ a_{(2)}(x)+\calo(p_1^4)\right)\,,
\end{split}
\eqlabel{pert}
\end{equation}
where the $p_1\to 0$ solution represents the UV $\cft$  de Sitter vacuum solution, and
following \eqref{bcdata}, the bulk scalar solution is normalized in the limit $x\to 0$ as 
\begin{equation}
p_{(1)}=x^{d+1-\Delta}\biggl(1+\calo(x)\biggr)+x^\Delta \biggl(\calo(x^0)\biggr)\,.
\eqlabel{p1norm}
\end{equation}
Furthermore, the location of the AH is determined from \eqref{xah} as
\begin{equation}
\begin{split}
0=&\biggl[\frac{(x-1)^2}{2x^2}+\left(
\frac{(3x-1) (x-1)^2}{2x} s_{(2)}'
+\frac{(x-1)^2}{2x^2} s_{(2)}+\frac{1-3x}{x} a_{(2)}
\right) 
p_1^2\\
&+\calo(p_1^4)\biggr]\bigg|_{x=x_{AH}}\,.
\end{split}
\eqlabel{ahfinder}
\end{equation}
Note that to leading order in $p_1$, $x_{AH}=1$ (which is the motivation for a
diffeomorphism condition \eqref{lstat}). 
Solving for $p_{(1)}$ in \eqref{pert} subject to \eqref{p1norm} results in
\begin{equation}
p_{(1)}=\frac{2^{1-\Delta}\sqrt{\pi}\Gamma\left(\Delta\right)}
{\Gamma\left(\Delta-\frac d2-\frac12\right)\Gamma\left(1+\frac d2\right)}\ _2F_1\biggl(
\Delta\,,\, d+1-\Delta\,;\, 1+\frac d2\,;\, \frac{3x-1}{2x}\biggr)\,.
\eqlabel{p1sol}
\end{equation}
As $z\equiv 1-x\to 0_+$ the bulk scalar solution diverges,
\begin{equation}
p_{(1)}=c_{sing}\ z^{-\frac d2}\left(1+\calo(z)\right)\,,\ c_{sing}=\frac{2^{\frac d2+1-\Delta}\pi^{3/2}}
{\Gamma\left(\Delta-\frac d2-\frac 12\right)
\Gamma\left(1-\frac d2\right)\Gamma\left(d+1-\Delta\right)\sin\frac{\pi d}{2}}\,.
\eqlabel{p1s}
\end{equation}
While there is no simple closed form solution for $s_{(2)}$ and $a_{(2)}$,
it is straightforward to extract their (singular) asymptotes as $z\to 0_+$,
\begin{equation}
s_{(2)}=z^{-d}\biggl(-\frac{c_{sing}^2}{8(d-1)}+\calo(z)\biggr)\,,\qquad
a_{(2)}=z^{-d}\biggl(-\frac{c_{sing}^2d}{8(d^2-1)}\ z+\calo(z^2)\biggr)\,.
\eqlabel{s2a2}
\end{equation}
Asymptotes \eqref{p1s} and \eqref{s2a2} are enough to compute the first subleading
correction to the location of the AH from \eqref{ahfinder}:
\begin{equation}
z_{AH}^{d+1}\equiv \left(1-x_{AH}\right)^{d+1}=\frac{c_{sing}^2d}{4(d+1)}\ p_1^2
\cdot \biggl(1+\calo\left(p_1^\frac{2}{d+1}\right)\biggr)+\calo(p_1^4)\,.
\end{equation}
Finally, we can compute the leading in the conformal symmetry breaking parameter $p_1$ (see \eqref{defp1}) VEE density
$s_{ent}$:
\begin{equation}
\frac{\kappa^2}{2\pi}\ \frac{s_{ent}}{H^d}=\left[\frac{\sigma_v(r)}{H}\right]^d\bigg|_{r=r_{AH}}=
\left[\frac{f(x)}{x}\right]^d\bigg|_{x=x_{AH}}=z_{AH}^d\biggl[1+\calo\left(p_1^{\frac{2}{d+1}}\right)\biggr]\,,
\eqlabel{send}
\end{equation}
or
\begin{equation}
\frac{\kappa^2}{2\pi}\ \frac{s_{ent}}{H^d}=\left(\frac{c_{sing}^2d}{4(d+1)}\right)^{\frac{d}{d+1}}
\cdot p_1^{\frac{2d}{d+1}}\cdot \biggl[1+\calo\left(p_1^{\frac{2}{d+1}}\right)\biggr]\ \propto\
\left(\frac{\Lambda}{H}\right)^{\frac{2d(d+1-\Delta)}{d+1}}\,.
\eqlabel{sent2}
\end{equation}
with $c_{sing}$ given by \eqref{p1s}.

The computations leading to \eqref{sent2} are valid when $\Delta$ is strictly above the 
Breitenlohner-Freedman bound, $\Delta>\Delta^{BF}\equiv \frac d2+\frac 12$. It is straightforward
to extend the analysis when $\Delta=\Delta^{BL}$ --- the only modification in \eqref{sent2}
is a different from \eqref{p1s} expression for $c_{sing}\equiv c_{sing}^{BL}$, 
\begin{equation}
c_{sing}^{BF}=-\frac{\pi^{3/2}}{\sqrt{2}\Gamma\left(1-\frac d2\right)\Gamma\left(\frac 12+\frac d2
\right)\sin\frac{\pi d}{2}}\,.
\eqlabel{csingbf}
\end{equation}

\begin{figure}[h]
\begin{center}
\psfrag{p}[c]{{$p_1=\frac{\Lambda^2}{H^2}$}}
\psfrag{q}[c]{{$p_1=\frac{\Lambda}{H}$}}
\psfrag{a}[ct]{{$\kappa^2 s_{ent}/(2\pi H^3)$}}
\psfrag{h}[cb]{{$\kappa^2 s_{ent}/(2\pi H^3)$}}
  \includegraphics[width=3.0in]{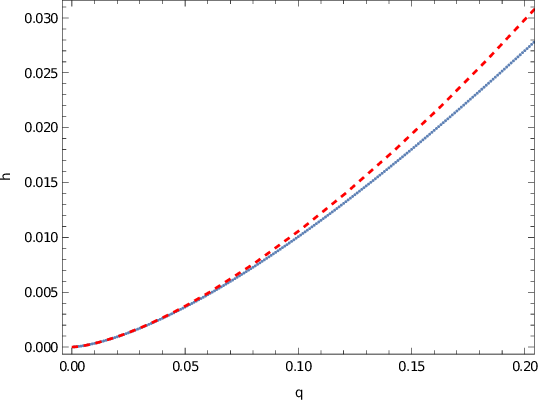}
  \includegraphics[width=3.0in]{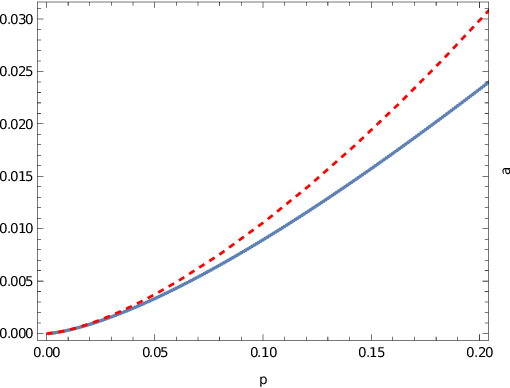}
\end{center}
 \caption{Vacuum entanglement entropy density $s_{ent}$ of de Sitter
 DFPs of $(d,\Delta)=(3,3)$ model (the left panel), and
 $(d,\Delta)=(3,2)$ model (the right panel).
 Solid curves are numerical fully non-linear in $p_1$ (see \eqref{defp1})
 computations of $s_{ent}$; the dashed red curves represent the
leading order  analytic approximations \eqref{s33} and \eqref{s32} valid
when $H\gg \Lambda$.
}\label{figure1}
\end{figure}

Consider now special cases of the general formula \eqref{sent2}.
\begin{itemize}
\item $(d,\Delta)=(2,2)$:
\begin{equation}
\frac{\kappa^2}{2\pi}\ \frac{s_{ent}}{H^2}=\frac{1}{6^{2/3}}\cdot p_1^{4/3}\cdot
\biggl[1+\calo\left(p_1^{\frac{2}{3}}\right)\biggr]\ \propto\ \left(\frac{\Lambda}{H}\right)^{\frac 43}\,,
\eqlabel{s22}
\end{equation}
reproducing the result of \cite{Buchel:2017lhu}.
\item $(d,\Delta)=(3,3)$:
\begin{equation}
\frac{\kappa^2}{2\pi}\ \frac{s_{ent}}{H^3}=\frac{3^{3/4}\pi^{3/2}}{2^{21/4}}\cdot p_1^{3/2}\cdot
\biggl[1+\calo\left(p_1^{\frac{1}{2}}\right)\biggr]\ \propto\ \left(\frac{\Lambda}{H}\right)^{\frac 32}\,.
\eqlabel{s33}
\end{equation}
\item $(d,\Delta)=(3,2)$:
\begin{equation}
\frac{\kappa^2}{2\pi}\ \frac{s_{ent}}{H^3}=\frac{3^{3/4}\pi^{3/2}}{2^{21/4}}\cdot p_1^{3/2}\cdot
\biggl[1+\calo\left(p_1^{\frac{1}{2}}\right)\biggr]\ \propto\ \left(\frac{\Lambda}{H}\right)^{3}\,.
\eqlabel{s32}
\end{equation}
\end{itemize}
Note that while to leading order in the conformal symmetry breaking parameter $p_1$
\eqref{s33} and \eqref{s32} are identical when expressed in $p_1$,
however, because of \eqref{defp1}, the $\frac{\Lambda}{H}$ scaling of $s_{ent}$ is different
for $\Delta=3$ and $\Delta=2$. In fig.~\ref{figure1} we test
predictions \eqref{s33} and \eqref{s32} against the fully nonlinear in $p_1$
computations of $s_{ent}$ in these two models.

\subsection{Effective temperature $T_{DFP}$}

Following the general arguments in section \ref{intro}, the maximal reheating temperature of a $\qft$
from the rapid exit of the prolonged inflationary stage is at least as high as
the effective DFP temperature $T_{DFP}$, computed from
\begin{equation}
\frac{s_{eq}(T)}{\Lambda^d}\bigg|_{T=T_{DFP}}=\frac{s_{ent}(H)}{\Lambda^d}\,,
\eqlabel{tdfpdef}
\end{equation}
where $s_{eq}$ is the thermal equilibrium entropy density of the theory in Minkowski spacetime,
and $s_{ent}$ is the VEE density of the inflationary phase -- in the case of prolonged inflation
a de Sitter DFP of the theory.
Note that we compare the dimensionless quantities in \eqref{tdfpdef}.
$s_{eq}$ can be computed from the thermal equation of state of the $\qft$, and since our theory is
massive, $s_{eq}$ will depend both on $T$ and $\Lambda$. For an efficient reheating
from the phenomenological perspective we
need
\begin{equation}
T_{DFP}\gg \Lambda\,,
\eqlabel{effreh}
\end{equation}
\ie  upon the inflationary exit, the de Sitter DFP state of the theory is reheated to a much higher
temperature than any mass scale of the theory --- for a QCD this would a reheating way above the
confinement scale of the theory.  From the entropy perspective, effective reheating implies
that
\begin{equation}
\frac{\kappa^2}{2\pi}\ \frac{s_{ent}(H)}{\Lambda^d}\ \gg 1\,.
\eqlabel{effent}
\end{equation}
Remarkably, the condition \eqref{effent} implies that the de Sitter DFP must be
in the regime of perturbatively small conformal symmetry breaking parameter $p_1$ \eqref{defp1}
of the $\qft$. Indeed, from \eqref{sent2} we
have
\begin{equation}
\frac{\kappa^2}{2\pi}\ \frac{s_{ent}(H)}{\Lambda^d}=\frac{\kappa^2}{2\pi}\ \frac{s_{ent}(H)}{H^d}
\cdot \left(\frac{H}{\Lambda}\right)^d\ \propto\ \left(\frac{\Lambda}{H}\right)^{\frac{2d(d+1-\Delta)}{d+1}}
\cdot \left(\frac{H}{\Lambda}\right)^d=\left(\frac{H}{\Lambda}\right)^{\frac{d(2\Delta-d-1)}{d+1}}\,,
\eqlabel{pertpr}
\end{equation}
so that \eqref{effent} is indeed true when $H\gg \Lambda$ and $\Delta>\Delta^{BL}=\frac 12+\frac d2$.

In the regime \eqref{effreh}, the thermal equation of
state of a $\qft$ is that of the UV $\cft$.
The thermal entropy density of the corresponding holographic $\cft$ can be easily computed,
we find
\begin{equation}
\frac{\kappa^2}{2\pi}\ \frac{s_{eq}^{CFT}}{T^d}=\left(\frac{4\pi}{d+1}\right)^d\qquad
\Longrightarrow\qquad s_{eq}=s_{eq}^{CFT} \biggl(1+\calo\left(\left(\frac{\Lambda}{T}\right)^{2(d+1-\Delta)}
\right)\biggr)\,.
\eqlabel{thermal}
\end{equation}
Using \eqref{sent2} and \eqref{thermal} we find from \eqref{tdfpdef}
the lower bound reported in \eqref{ltdfp},
\begin{equation}
\frac{T_{DFP}}{\Lambda}=\frac{d+1}{4\pi}\cdot \biggl(\frac{c_{sing}^2d}{4(d+1)}\biggr)^{\frac{1}{d+1}}\cdot
\left(\frac{H}{\Lambda}\right)^{\frac{2\Delta-d-1}{d+1}}\ \gg\ 1\qquad {\rm when}
\qquad H\gg \Lambda\,,
\eqlabel{fint}
\end{equation}
for $\Delta>\frac 12+\frac d2$.
Consider now some special cases of the general formula \eqref{fint}.
\begin{itemize}
\item $(d,\Delta)=(2,2)$:
\begin{equation}
\frac{T_{DFP}}{\Lambda}=\frac{3^{2/3}}{2^{7/3}\pi}\cdot\left(\frac{H}{\Lambda}\right)^{\frac 13}\qquad {\rm when }\qquad H\gg \Lambda\,.
\eqlabel{t22}
\end{equation}
\item $(d,\Delta)=(3,3)$:
\begin{equation}
\frac{T_{DFP}}{\Lambda}=\frac{3^{1/4}}{2^{7/4}\sqrt{\pi}}\cdot\left(\frac{H}{\Lambda}\right)^{\frac 12}\qquad {\rm when }\qquad H\gg \Lambda\,.
\eqlabel{t33}
\end{equation}
\end{itemize}

\begin{figure}[h]
\begin{center}
\psfrag{z}[c]{{$\frac{\Lambda}{H}$}}
\psfrag{w}[c]{{${T}/{\Lambda}$}}
\psfrag{y}[ct]{{$\kappa^2 s/(2\pi H^3)$}}
  \includegraphics[width=3.0in]{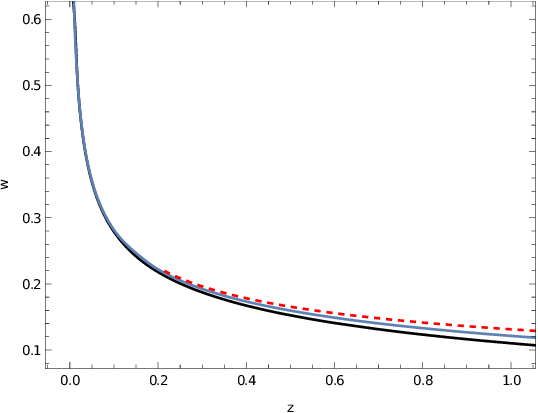}
 \includegraphics[width=3.0in]{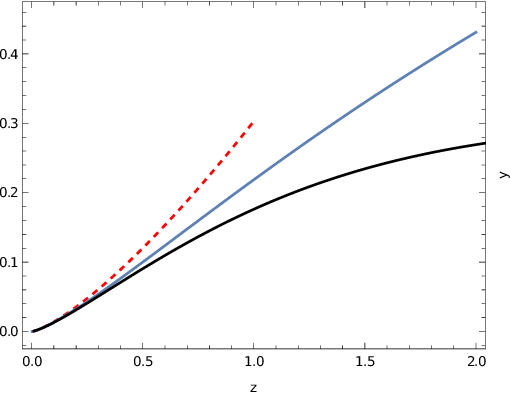}
\end{center}
 \caption{(The left panel) The maximal reheating temperature $T_{HBB}$ (the blue curve,
 from \cite{Buchel:2023djl})
 and the effective temperature $T_{DFP}$ (the black curve) of the de Sitter dynamical fixed point of
$(d,\Delta)=(2,2)$ model as a function of the conformal symmetry breaking parameter. (The right panel)
The maximal thermal entropy density of the Hot Big Bang
$s_{HBB}$ (the blue curve,  from \cite{Buchel:2023djl}) and the VEE density of the
de Sitter DFP $s_{ent}$ (the black curve) in this model. The red dashed curves represent the leading approximations
for $H\gg \Lambda$.
}\label{figure2}
\end{figure}

We conclude this section with the conjecture that $T_{DFP}$ computed in the regime $H\gg \Lambda$
is actually the maximal reheating temperature $T_{HBB}$ itself, \ie the bound in \eqref{lbt}
is saturated.  The reason being is that for an efficient reheating, the de Sitter DFP
state of the theory must be very entropic from the $\qft$ perspective, see \eqref{effent}.
Thus, following the ideas of the ``eigenvalue thermalization hypothesis''
\cite{Srednicki:1994mfb} it must be to a good approximation
thermal. This conjecture passes the test of the full-fledged simulations of 
$(d,\Delta)=(2,2)$ model discussed in \cite{Buchel:2023djl}.
In fig.~\ref{figure2} we compare the temperatures (the left panel)
and the entropy densities  (the right panel) in the model
at de Sitter DFPs (the black curves)
and upon thermalization following the inflationary exit (the blue curves).
The dashed red curves represent the leading approximation in the small conformal
symmetry breaking parameter \eqref{defp1} given by \eqref{t22} for the
temperature and \eqref{s22} for the entropy density.
Computations of the maximal reheating temperature $T_{HBB}$ and
the Hot Big Bang thermal entropy density $s_{HBB}$
(the blue curves in fig.~\ref{figure2})
require 
holographic simulations of the rapid, \ie
in the regime $\gamma\gg H$,
inflationary exit \eqref{exita} of the corresponding model.
For  $(d,\Delta)=(2,2)$ model this was done in
\cite{Buchel:2023djl}\footnote{$T_{HBB}$ is referred to as $T_r^{max}$ in
\cite{Buchel:2023djl}. $s_{HBB}$ is the same as the equilibrium thermal
entropy density of the model in Minkowski space-time evaluated
at the temperature $T=T_r^{max}$.}.
Notice that upon thermalization,
the equilibrium entropy density is always larger than the corresponding VEE $s_{ent}$
of the de Sitter DFP, however, the entropy production during the thermalization
process vanishes as $\frac\Lambda H\to 0$. Correspondingly, the maximal Hot Big Bang reheating
temperature $T_{HBB}$ is always larger than $T_{DFP}$, however, the difference between the
two also vanishes as $\frac\Lambda H\to 0$.

\section{Reheating in $\caln=2^*$ gauge theory}
\label{n2}

\begin{figure}[h]
\begin{center}
\psfrag{x}[c]{{$\frac{m}{H}$}}
\psfrag{y}[c]{{${T_{DFP}}/{m}$}}
  \includegraphics[width=4.0in]{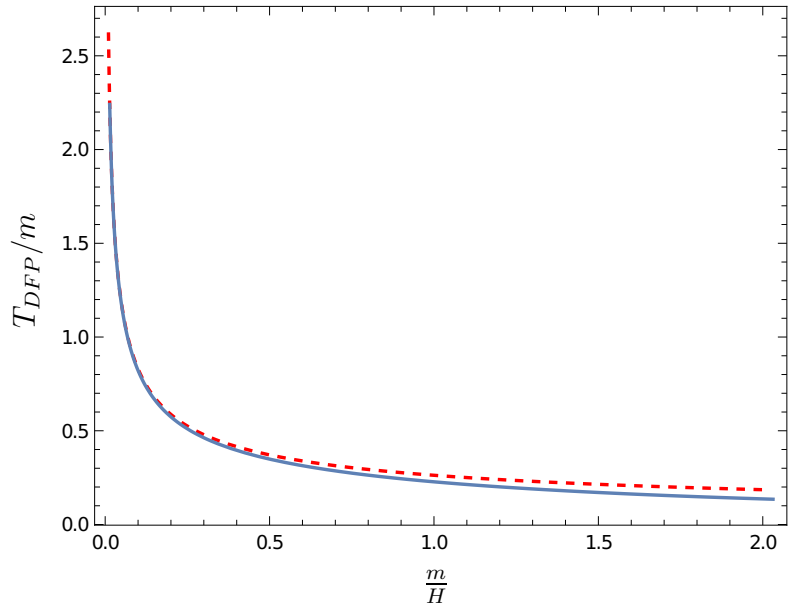}
\end{center}
 \caption{The lower bound (the blue curve) on the maximal reheating
 temperature $T_{DFP}$ of $\caln=2^*$ gauge theory
 in the rapid exit from the prolonged inflationary stage. The dashed red curve
 corresponds to small conformal symmetry breaking parameter, \ie for $H\gg m$, in
 $(d,\Delta)=(3,3)$ model, see \eqref{t33}. 
}\label{figure3}
\end{figure}

The massive holographic models discussed in section \ref{perd} involved a
deformation of a UV $\cft$ by a single relevant operator $\calo_\Delta$.
More common are holographic models where multiple relevant operators are sources ---
\eg the four ($d=3$)
space-time  dimensional $\caln=2^*$ gauge theory \cite{Pilch:2000fu,Buchel:2000cn}. Here, there
are two relevant operators $\{\calo_2,\calo_3\}$ with the sources being
the masses of the bosonic $p_1^{\Delta=2}\propto m_b^2$ and the
fermionic $p_1^{\Delta=3} \propto m_f$ components of the $\caln=2$ hypermultiplet.
For generic values of $m_b\ne m_f$ the supersymmetry in the model is completely broken.
We consider\footnote{There is a consistent
truncation of the $\caln=2^*$ holographic model to $m_f=0$ and $m_b\ne 0$.
However, in this case the deformation operator is $\calo_2$ with
$\Delta=2=\Delta^{BF}$, and thus it can not lead to an efficient
reheating.} the $\caln=2$ supersymmetry preserving masses, $m_b=m_f\equiv m$.  
de Sitter DFPs of the model were studied in \cite{Buchel:2017pto,Buchel:2019qcq}
and Minkowski space-time thermodynamics of the model was analyzed in \cite{Buchel:2007vy}.
Using the results from these papers, and the correspondence relation \eqref{tdfpdef},
we produce in fig.~\ref{figure3} the maximal reheating temperature in $\caln=2^*$
gauge theory (the blue curve). The dashed red curve represents the leading approximation  
to $T_{DFP}$ in the limit of small conformal symmetry breaking $\frac{m}{H}\to 0$,
see\footnote{The  $\frac 12$ scaling exponent is correct, but the prefactor
must be modified to account for the fact that the bulk scalar dual to $\calo_3$
does not have a canonical kinetic term; furthermore, the precision holography
in this case requires that $m\propto \Lambda$, but is not equal to it
\cite{Buchel:2000cn,Buchel:2013id}.
} \eqref{t33}.
This example illustrates that the regime of the efficient reheating is controlled
by the relevant operator with the largest scaling dimension.

\section{Reheating in the cascading gauge theory}
\label{ksr}

The general formula \eqref{fint} implies that the most efficient reheating occurs in theories with the
largest dimension sourced operator $\calo_\Delta$. Since renormalizability of a $\qft$ requires that
sources are introduced only for marginal and/or relevant operators, and exactly marginal operators do
not reheat \eqref{fmag}, we look for a model with a marginal, but not exactly marginal sourced operator.
An example of just such a model is the $d=3$, $\caln=1$ supersymmetric
$SU(N+M)\times SU(N)$ Klebanov-Strassler cascading gauge theory
\cite{Klebanov:2000hb,Herzog:2001xk,Aharony:2005zr,Buchel:2021yay}. de Sitter DFPs of
the cascading gauge theory were studied extensively in \cite{Buchel:2019pjb}; here,
\begin{equation}
\frac{s_{ent}}{H^3}\  \propto\ M^4 \left(\ln\frac{H^2}{\Lambda^2}\right)^{-3/4}\qquad
{\rm as }\qquad H\gg \Lambda\,,
\eqlabel{sentcas}
\end{equation}
with $\Lambda$ being the strong coupling scale of the theory \cite{Buchel:2021yay}.
The high-temperature Minkowski space-time thermodynamics of the theory was analyzed in 
\cite{Buchel:2000ch,Aharony:2007vg},
\begin{equation}
\frac{s_{eq}}{T^3}\  \propto\ M^4 \left(\ln\frac{T^2}{\Lambda^2}\right)^{2}\qquad
{\rm as }\qquad T\gg \Lambda\,.
\eqlabel{seqcas}
\end{equation}
From \eqref{tdfpdef} we estimate the lower bound on the maximal reheating $T_{DFP}$
as in \eqref{tdfpkt}.

\section*{Acknowledgments}
I would like to thanks Iosif Bena, Elias Kiritsis and Giuseppe Policastro
for valuable discussions. I am grateful to The Institute of Theoretical Physics (IPhT)
in Saclay for hospitality where part of this work was done.
Research at Perimeter Institute is supported in part by the Government
of Canada through the Department of Innovation, Science and Economic
Development Canada and by the Province of Ontario through the Ministry
of Colleges and Universities. This work is further supported by a
Discovery Grant from the Natural Sciences and Engineering Research
Council of Canada.

%\appendix

\bibliographystyle{JHEP}
\bibliography{holreheat}

\end{document}